\documentclass[a4paper,12pt]{article}
\usepackage[numbers]{natbib}
\usepackage{amsmath}
\usepackage{here}
\usepackage{amsthm}
\usepackage{graphicx,color}
\usepackage{ascmac}
\usepackage{amsfonts}
\usepackage{amssymb}
\usepackage{authblk}
\usepackage{graphicx}
\usepackage{tikz}
\usetikzlibrary{intersections,calc,arrows.meta,math}
\usepackage{fancybox}
\usepackage{tikz-qtree}
\usepackage{latexsym}

\usepackage{breqn}
\usepackage{mathtools}
\usepackage{subfigure}
\newtheorem{assumption}{\bf Assumption}
\newtheorem{proposition}{\bf Proposition}
\newtheorem{lemma}{\bf Lemma}

\begin{document}
	\title{Shifting to telework and the firm's location choice: Does telework make our society efficient? \thanks{I am grateful for Ryo Itoh, Tomoya Mori, Se-il Mun, Sugie Lee, Jacques-François Thisse, Tomohiro Machikita, Minoru Osawa, Mitsuru Ota, Tatsuhito Kono, Dao-Zhi Zeng, Naoya Fujiwara, and Tomokatsu Onaga gave me useful and helpful comments. The preliminary version of this study was presented in Mathematical geographical modelling for environmental humanities in Kyoto Univarsity, Urban Economics Workshop in Kyoto University,  11th European Meeting of the Urban Economics Association (UEA) in London, and 35th The Applied Regional Science Conference (ARSC). This work was supported by JST the establishment of university fellowships towards the creation of science technology innovation, Grant Number JPMJFS2102, JST SPRING, Grant Number JPMJSP2114, the Research Institute for Mathematical Sciences, an International Joint Usage/Research Center located in Kyoto University, and Research funds for student-initiated projects in Tohoku university. E-mail: \textcolor{blue}{\tt{tsuboi@se.is.tohoku.ac.jp}} }}
	\author[]{Kazufumi Tsuboi $^1$}
	\affil[]{1. \it{Graduate School of Information Sciences, Tohoku University, Aoba 6-3-09, Aramaki, Aoba-ku, Sendai, Miyagi 980-8579, Japan.}}
	\date{}
	\maketitle
	\begin{abstract}
Although it has been suggested that the shift from on-site work to telework will change the city structure, the mechanism of this change is not clear. This study clarifies how the location of firms changes when the cost of teleworking decreases and how this affects the urban economy. The two main results obtained are as follows. (i) The expansion of teleworking causes firms to be located closer to urban centers or closer to urban fringes. (ii) Teleworking makes urban production more efficient and cities more compact. This is the first paper to show that two empirical studies can be represented in a unified theoretical model and that existing studies obtained by simulation can be explained analytically.
	\end{abstract}
\newpage
\section{Introduction}	
\begin{figure} [H]
	\centering
	\includegraphics[width=0.6\linewidth]{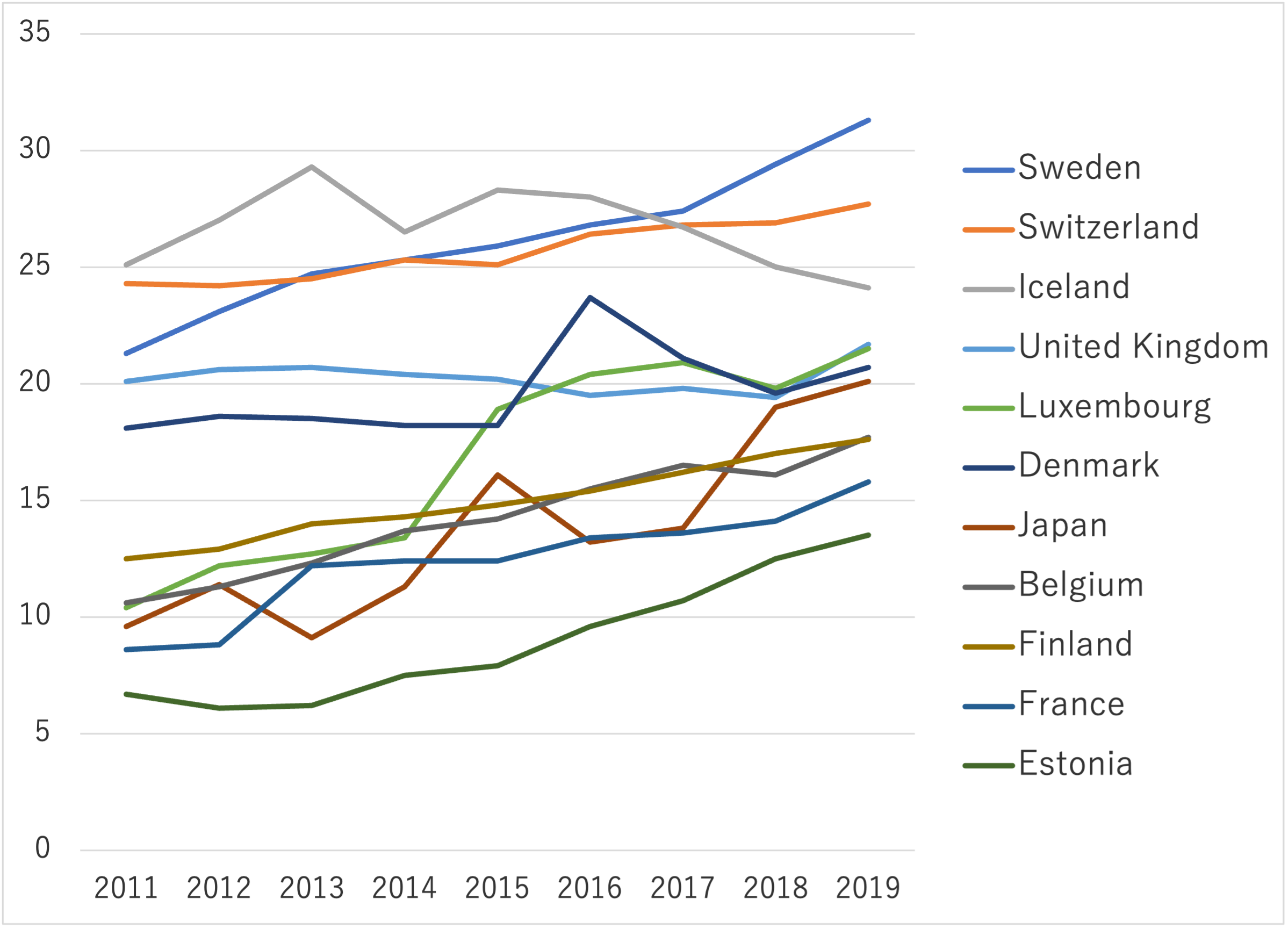}
	\caption{\textbf{The percentage of teleworkers employed by one firm in EU and Japan :} This figure shows the percentage of teleworkers employed by one firm. The percentage of teleworkers increased each year.}
	\label{TeleTre}
\end{figure}

	Recently, technological innovation has led to the expansion of telework around the world.
	New application tools such as Skype, Zoom, and Slack have made teleworking possible at lower costs, which has rapidly replaced office work with telework.
	Looking at EU and Japanese data in Figure~\ref{TeleTre}, the percentage of teleworkers employed by firms was on the rise, even if we focus on the years 2011-2019 before COVID-19. 
	\footnote{Data are from \textit{Communications Usage Trend Survey (Tu-shin Riyou Doukou Chiyousa)} published by the Ministry of Internal Affairs and Communications and \textit{Employed persons working from home as a percentage of the total employment, by sex, age and professional status} published by Eurostat}

	Although it has been pointed out that the shift form on-site work to telework will change the location of households and firms \cite{nilles1975telecommunications}, the mechanism for the change in the location of firms is not clear.
	Households have less opportunities to commute to their office and move out from city center to suburban ~\cite{de2018working}.
	Firms decrease their floor space as fewer workers commute to their offices and change their locations.
	However, there is an ongoing debate about where firms employing teleworkers should be located in the city.
	Existing empirical studies have shown both the results of firms employing teleworkers agglomerated to the city center and9 dispersed to the suburbs.
	In Maeng and Nedovic-Budic (2010)~\cite{maeng2010relationship}, IT industries that are more likely to telework in Washington are agglomerated to the city center.
	In Liao (2012)~\cite{liao2012inshoring}, firms with high rates of telework implementation (e.g., call centers and IT industry) disperse to the suburbs.
	Moreover, existing theoretical models cannot explain this phenomenon and it remains an unsolved puzzle because they set all firms to telework at the same level, such as lockdown during COVID-19.
	However, the location of firms employing teleworkers is an important issue not only for explaining empirical studies, but also for analyzing the social impact of telework policies.
	
	Hence, this study aims to clarify how the location of firms changes when the cost of teleworking decreases and how this affects the urban economy.
	Our analysis consists of two issues.
	First, we analyze the location of firms employing teleworkers.
	Second, we explain how the expansion of teleworking changes economic activity, such as wages, welfare, city size, and land rents.
	
	In this paper, we construct a model that endogenously determines the location of household and two types of firms with different rates of telework implementation based on Ogawa and Fujita (1980)~\cite{ogawa1980equilibrium}.
	Of the two types of firms, one is \textit{an office firm} which hires only office workers and the other is \textit{a telework firm} which hires both teleworkers and office workers.
	We identify the location of firms that conduct telework in a setting where the cost of teleworking decreases due to technological innovation using bid rent functions.
	We also conduct comparative statics for each spacial equilibrium to analyze the effects of the spreading telework on the size of cities, land rents, and welfare.
	
	Many researchers have examined the relationship between telework and urban economies.
	Most theoretical studies using simulations have concluded that business districts shrink, land rents and traffic congestion in urban centers decrease, and welfare increases by teleworking \cite{safirova2002telecommuting}, \cite{rhee2008home}, and \cite{rhee2009telecommuting}.
	Delventhal et al. (2021)~\cite{delventhal2022jue} and Lennox (2020)~\cite{lennox2020more} showed that when teleworking expanded rapidly due to COVID-19, households moved to the suburbs and firms are agglomerated in the urban center.
	
	Kyriakopoulou and Picard (2021)~\cite{kyriakopoulou2022zoom} is the most relevant to our study and this study supports their results; Kyriakopoulou and Picard (2021)~\cite{kyriakopoulou2022zoom} identifies the location of households and firms when all firms conduct telework at the same level.
	They find that teleworking reduced the size of business districts, decreased land rents near urban centers, and when the number of firms and households was large, increased intra-urban wages and social welfare, which is the same result as the comparative statics conducted by this study,
	
	The unique feature of this study is that two types of firms with different rates of teleworking select endogenously their location.
	In previous studies~\cite{safirova2002telecommuting}, \cite{rhee2008home},  \cite{rhee2009telecommuting}, and \cite{kyriakopoulou2022zoom}, all firms conduct telework at the same level.
	However, empirical studies indicated that there are differences in telework implementation rates by industry\cite{dingel2020many}, and that location also varies with telework implementation rates~\cite{liao2012inshoring}.
	Hence, in this study, we constructed and analyzed a model in which firms were divided into two types according to their telework implementation rates and endogenously select their locations.
	
	This paper introduces two innovative results and implications for the following issues.
	The first issue is to analytically clarify the location of firms employing teleworkers.
	Such firms are located in the boundary areas between households and firms (\textit{CBD fringes}) or between households and agricultural land (\textit{urban fringes}) according to several parameters: face-to-fave communication cost $ \tau $, commuting cost $ \kappa $, and teleworker ration $ \beta_t $ which is the percentage of employees who telework in the firm.
	If both $ \kappa/\tau $ and $ \beta_t $ are high (low), telework firms are located at \textit{urban fringes} (\textit{CBD fringes}).
	This is the first paper to  demonstrate two empirical results, \cite{maeng2010relationship} and \cite{liao2012inshoring}, in a unified theoretical model and to show the difference depends on the several key parameters.
	
	The second issue is to indicate that expanding telework downsizes cities and increases firm productivity, wages, and welfare by comparative statics.
	This result shows that existing studies obtained by simulation can be explained analytically and supports Kyriakopoulou and Picard (2021)~\cite{kyriakopoulou2022zoom} is not an analysis limited to short-term lockdowns.
	
	This paper is organized as follows.
	Section~\ref{Section2} proposes a model where I define the individual behavior of firms and households.
	Section~\ref{Section3} shows some equilibria and get the bid rent functions.
	Section~\ref{Section4} proposes spatial equilibrium patterns.
	Section~\ref{Section5} focus on telework firms' input and their location.
	Section~\ref{Section6} remarks the location externalities of telework firms to other economic sectors. 
	Section~\ref{Section7} states the conclusion of this paper.

\section{Model}\label{Section2}
We consider a linear city with two types of firms (office firms and telework firms), households, and absentee landlords.
The city expands on the unit-width segment where firms and workers interact through competitive labor and land markets.
There are two forces that promote the formation of business and residential areas: business face-to-face communication by on-site workers and workers' commuting costs.
Firms are located close to each other for saving face-to-face communication costs.
We follow the literature by considering agglomeration benefits through professional interaction and exchange of information across on-site workers.
As in some previous urban economics model, households face a trade off: locating far from firms incurs their high commuting cost but decreases their land rents.
The balance of those forces determines the land use pattern.

Firms and workers compete for land at a location $ y $ and $ r $ ($ y, r \in \mathbb{R} $). 
We assume a closed cities where a mass of $N$ households and a mass of $M$ firms that produce in the city are given exogenously.
The land is assumed tobe owned by absentee landlords following previous studies~\cite{ogawa1980equilibrium},~\cite{alonso2013location}, and~\cite{fujita1982multiple}.
\subsection{Two types of firms}
There are $M$ firms, each producing some kind of services or information which is exported to the national market at a constant price $p$.
We consider two types of firms: one is \textit{an office firm} (type $o$) which hires only office workers and the other is \textit{a telework firm} (type $t$) which hires both teleworkers and office workers.
The production function of each type $i\in \left\{o,t \right\}$ firm is of an input-output type, whose inputs are office worker, teleworker, and land, given by
\begin{equation}
	q_i=\min \{ \frac{L_{o,i}}{ (1-\beta_i)a_L}, \frac{L_{t,i}}{\beta_i a_L}  ,\frac{S_i}{a_{s,i}}\} ,
	\label{production function}
\end{equation}
where $ q_i =$ the amount of output produced, $S_i=$ land, $ L_{o,i} =$ office worker in type $i$ firm, $ L_{t,i} =$ teleworker in type $i$ firm, $ a_{s, i} =$  the land/output ration
\footnote{The bigger $ a_{s, i} $ is, the less land productivity the firm has because such a firm needs more land inputs.}
, $a_L =$ the labor/output ratio, and $\beta_i =$ teleworker ratio in the firm ($\beta_i \in [0,1]$) 

Furthermore, for simplicity, we assume that each firm produces the assume positive constant amount of output $\bar{q}=1$ and $ a_L=1 $.
\footnote{Although this assumption may seem too strong, $\bar{q}$ and $ a_L $ are irrelevant to the result. 
	$ \bar{q}_i $ disappear when we calculate bid rent functions.}
Consequently, from (\ref{production function}) each firm requires
\begin{equation}
	\bar{L}_{o,i}=(1-\beta_i)a_L\bar{q}_i,\hspace{10pt}\bar{L}_{t,i}=\beta_i a_L\bar{q}_i,\hspace{10pt} \bar{S}_i=a_{s, i}\bar{q}_i,
	\label{input function}
\end{equation}

For simplicity, we assume that \textit{telework firms} hire both teleworkers and on-site workers $ \beta_t \in (0,1) $ and \textit{office firms} hire only on-site workers $ \beta_o =0 $.
And \textit{office firms} are assumed to require more lands to produce goods than \textit{telework firms}.

Another important assumption for firms is that production requires face-to-face (FTF) communications among on-site workers.
We introduce two additional assumptions following previous research, \cite{ogawa1980equilibrium}, \cite{kyriakopoulou2022zoom}, and~\cite{fujita1996economics}.
(i) Each on-site worker communicates strictly equiprobably with every location of firms in the city.
(ii) Each communication is separately performed by some kind of communication method and the cost of the communication between any two firms is proportional to the distance between them, i.e. $ \tau|y-x| $ where $ \tau $ is the communication cost per unit distance.

Under these assumptions, on-site worker in a firm at $y$ will communicate with every location of firms in the city, and its total FTF communication costs $T(y)$ are given by
\begin{equation*}
	T(y)= \tau (1-\beta_i) \int_{-f}^{f}|y-x|dx, \hspace{10pt} x\in\{x | m(x)>0\},
\end{equation*}
where $ m(x) =$ density function of firms at each location $x$.
\footnote{Many previous studies have written the communication costs as $T(y)=\tau \int_{-f}^{f} m(x)|y-x|dx$.
	However, in this study, $m(x)$ is excluded because there are two types of firms and the calculation becomes complicated.  Even if $ m(x) $ is added, the qualitative conclusion remains the same.}
Using (\ref{production function}) and (\ref{input function}), profits of a firm locating at $y$ are given by
\begin{equation}
	\begin{split}
		\pi_i(y)=\underbrace{p\bar{q}_i}_{Revenue}&-\underbrace{\beta_i\bar{q}_i(w_{t}+MC_t)-(1-\beta_i)\bar{q}_iW(y)}_{Wages}\\
		&-\underbrace{a_{s,i}\bar{q}_i\phi_i(y)}_{Rent}-
		\underbrace{\tau(1-\beta_i)\bar{q}_iT(y)}_{Communication\,cost},
	\end{split}
	\label{Profit}
\end{equation}
where $MC_t$ represents telework costs and $ \phi_i(y) $ is the land rent of type $ i $ firm for a unit of land at $y$.
Thus, the objective of the firm is essentially to chose its location $y$ considering the distribution of all other firms in the city, so as to maximize its profits given by (\ref{Profit}).
Since all firms within the same type are assumed to be identical, the profit level of all firms within the same type must be the same at the equilibrium regardless of their locations.
Under free entry, profits are pushed to zero (zero profit condition): $\pi_i(y)=0$.


\subsection{Households}
There are $N$ identical households which have the same preferences.
The utility level $U$ of each household depends on the amount of land occupied (lot size) $h$ and the amount of composite good $z$.
Thus, the utility function is expressed by
\begin{equation}
	U(z,\bar{h}; r)=z .
	\label{Utility}
\end{equation}
The composite good is imported from the national market at a price normalized to 1.
Each household has one worker supplying its labor to a firm.
The income of the household is equal to the wage earned by that worker from that firm.
The only travel in the city is the journey to work and the commuting cost is proportional to the distance between the residence and the job site.
The commuting cost per unit distance $\kappa$ is assumed to be a given positive constant.
Thus, the budget constraint of a household locating at $r$ and working at $y$ is given by 
\begin{equation}
	\max\{W(y)-\kappa | r-y|, w_t\}= z(r)+ \psi(r)\bar h,
	\label{Budget constraint}
\end{equation}
where $W(y)$ is the wage of on-site workers and $w_t$ is the wage of teleworker paid by a firm locating at $y$, $| r-y|$ is the distance between the residence and the job site, $ \psi(r) $ is the land rent for a unit of land at $r$.
The object of each household is to maximize its utility (\ref{Utility}) subject to the budget constraint (\ref{Budget constraint}) by choosing $z, h, r,$ and $y$.

However, for the simplicity of analysis, in this article we consider only the case where the lot size of each household is fixed at some positive constant $\bar{h}$.
Accordingly, the objective of household is equivalent to choosing the residential location $r$ and the job site $y$ to maximize the amount of composite good,
\begin{equation}
	\max z= \max\{W(y)-\kappa| r-y|, w_t\}-\psi(r)\bar h
	\label{EQ.Hosueholds}
\end{equation}
Since all the households are assumed to be identical, all the households should achieve the same utility level $U^*$ and hence the same consumption of composite good $z^*$, where $U^*=U(z^*,\bar{h}$)
\section{Equilibrium conditions and bid land rents}\label{Section3}
Focusing on the market equilibrium of the model presented in Section~\ref{Section2}, we derive a bid rent function.
\subsection{Equilibrium conditions}
Having described   the behavior of households and firms, the rest of our task is to obtain the equilibrium solution for the following set of unknown functions and variables: (a) type $i$ firm density function $ m_i(x) $, (b) household density function $ n(x) $, (c) land rent profile $ R(x) $, (d) wage profile $ W(x) $, (e) commuting pattern of households locating at $r$ and commuting to job site $y$, $P(r,y)$, (f) utility level $ z^\ast $, and (g) the ratio of office firm $\theta$, for $ -f \leq x \leq f $, where $f$ and $-f$ are urban fringe distances.

Then, the necessary and sufficient conditions for these variables, $ m_i(x) $, $ n(x) $, $ R(x) $, $ z^\ast $, $\theta$ to represent an equilibrium land use pattern are summarized as follows.

\subsubsection{Land market} 
\begin{align}
	\begin{aligned}
		&R(x)=\max\{\psi(x),\phi_i(x),R_A\},\\
		&R(x)=\psi(x)\ if\ n(x)>0, \\
		&R(x)=\phi_i(x)\ if\ m_i(x)>0,\\
		&R(-f)=R(f)=R_A.
	\end{aligned}
	\label{EQ:Rent}
\end{align}
At each $x \in[-f,f]$,
\begin{equation}
	\bar{h} n(x)+\sum_{i\in \{o,t\}}a_{s,i} m_i(x)  +(land\ for\ agricultural\ use)=1.
	\label{EQ:Land}
\end{equation}
From (\ref{EQ:Rent}), each piece of land must be occupied by a household, a firm, or a farm which bids the highest rent at each location.
If households or firms are located at $x$ respectively, then they must have succeeded in bidding for land at that location.
The physical constraint on the amount of land are given by (\ref{EQ:Land}).
\subsubsection{Population constrain}
\begin{equation}
	\int m_o(x) dx=\theta M,\hspace{10pt}\int m_t(x) dx=(1-\theta) M, \hspace{10pt}\int n(x) dx=N,
	\label{EQ:Firm&PopulationNumber}
\end{equation}
where the ratio of office firms to telework firms $\theta$ is determined in spatial equilibrium.
We assure no unemployment in the city $M=N$.

\subsubsection{Labor market}
At each $x\in[-f,f]$,
\begin{equation}
	m_o(y)= \int n(r)P(r,y) dr,
	\quad
	(1-\beta_t)m_t(y)= \int n(r)P(r,y) dr,
	\label{EQ:On-siteDemandSupply}
\end{equation}
where the demand for on-site workers in office firm and telework firm must be equal to the supply of on-site workers at all locations in the city.
\begin{equation}
	\beta_t(1-\theta)M=\int n(r)P(r,r) dr,
	\label{EQ:teleDemandSupply}
\end{equation}
where the demand for teleworkers in telework firm must be equal to the supply of teleworkers.

\subsubsection{Commuting}
Following \cite{alonso2013location} and \cite{ogawa1980equilibrium}, a worker who works at $ y $ and resides at $ r $ is assumed to incur commuting costs by $ \kappa | y-r | $.
If she chose commute to $y$, the wage at $y$ is the highest salary in the city,
\begin{equation}
	W(y)-\kappa | y-r |= \max_{x} \{ W(x)+\kappa | x-r |\}.
	\label{CommutEQ}
\end{equation}
\subsection{Bid land rents}
From  (\ref{Profit})  and (\ref{EQ.Hosueholds}) , we can yield bid rent functions.
Firms at $ x $ have the same profit $ \pi^\ast (x) $ in equilibrium and households at $ x $ have the same indirect utility $ z^\ast (x) $. 
From (\ref{Profit}), the firm's bid rent of type $ i $ is given by
\footnote{The slope and curvature are
	\begin{equation*}
		\phi_i'(x)=\frac{1-\beta_i}{a_{s,i}}(W'(x)-\tau  T'(x))
		\gtrless 0, \quad \phi_i''(x)=-\frac{1-\beta_i}{a_{s,i}} \tau T''(x)< 0,
		\label{FirmRentSlope}
	\end{equation*}
	where different type of firm has different gradients and the curvatures.}
\begin{equation}
	\phi_i(x)=\frac{1}{a_{s_i}}\left \{ p-\beta_i(w_{t}+MC_t)-(1-\beta_i)W(x)-\tau  (1-\beta_i)T(x) \right \}.
	\label{RentFirm}
\end{equation}
From (\ref{EQ.Hosueholds}), the worker's bid rent is given by
\begin{equation}
	\psi(x)=\frac{1}{\bar h}\{W(y)-z^\ast\}.
	\label{RentHouseholds}
\end{equation} 
\section{The change in location equilibrium due to the emergence of firms employing teleworkers.}\label{Section4}
Consider a city where the high cost of telework gradually decreases with the development of ICTs.
The decrease in costs causes teleworking to expand and change the location equilibrium.
The appearance of firms in a city employing teleworkers will also affect economic activities such as wages and social welfare.

In this section, we analyze land use patterns of households and firms by equilibrium conditions and land rents (\ref{Profit})-(\ref{RentHouseholds}).
We will show a few land use patterns are equilibrium if the parameters satisfy certain conditions.
The equilibrium land use pattern determines city boundaries, {\it CBD fringes} $ b $  and {\it urban fringes} $ f $.
{\it CBD fringes} $ b_i (i=1, 2, 3...) $ are boundaries within the city.
{\it Urban fringes} $ f $ are boundaries between the city and agricultural land $ R_A $.
Following previous research \cite{ogawa1980equilibrium} and \cite{fujita1982multiple}, we define the center of the business district as $ x=0 $ and examine the right half of the city.

\subsection{Location equilibrium when teleworking firms are not in the city}
\begin{center}
	\begin{figure}[h]
		\centering
		\includegraphics[width=300pt]{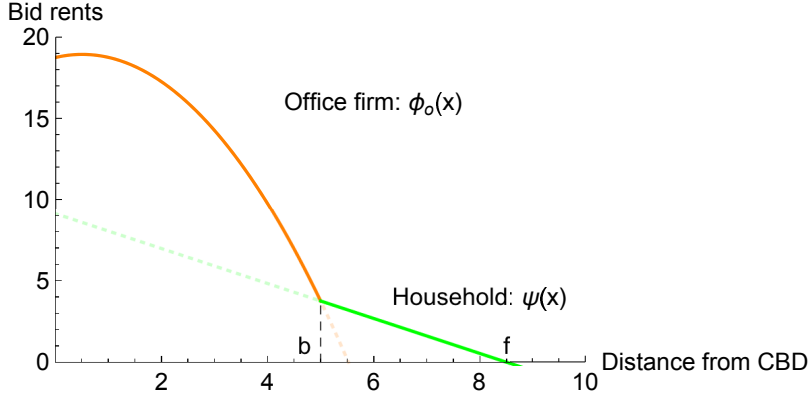}
		\caption{\textbf{An equilibrium land use pattern without telework: Completely segregation}
			The bid rent curve of office firm represents the orange line and that of household represents the green line.
			Benchmark parameter values are given by $ p=50, M=50,   \kappa=0.15, \tau=0.15, a_{s,o}=0.2, a_{s,t}=0.18, \bar{h}=0.14, R_A=0$. These benchmark parameter values are the same in the later analysis but we will change $ \beta_t $ and $ MC_t $.}
		\label{Benchmark}
	\end{figure}
\end{center}
Initially, consider cities where ICTs are underdeveloped and the cost of teleworking is high.
There, firms employing teleworkers do not appear in the city and the location equilibrium is essentially the same as in previous studies.

A monocentric city as the benchmark case contains a central business district on $ \left[ -b, b\right] $ and is surrounded by two symmetric residential districts on $ \left[ b, f\right] $ and $\left[ -f, -b\right] $ with $(0<b<f)$ as in Fig \ref{Benchmark}.
\footnote{Essentially, the benchmark case is the same of OF model.
	Completely and incompletely mixed land use pattern appears.}
A monocentric city without telework firms is set as the benchmark case.
When telework costs $MC_t$ is high, telework firms cannot appear in the city.
City boundaries are exogenously given by 
\footnote{Detailed calculation is in Appendix~\ref{BordersBench}.}
\begin{equation}
	b= \frac{a_{s,o}}{2} M,\quad
	f= \frac{\bar{h}+a_{s,o}}{2} M.
	\label{BoundariesBenchmark}
\end{equation}
From (\ref{CommutEQ}) and (\ref{BoundariesBenchmark}), FTF communication costs and wages are given by
\begin{equation}
	T(x)=x^2+b^2, \quad W(x)=w - \kappa x, \quad x\in\left[\,0,b \right),
	\label{FTFandWage}
\end{equation}
where $ w $ is the wage paid by firms at $ x=0 $ which is determined endogenously.
At city boundaries $b$ and $f$, bid rents satisfy the following necessary conditions for spatial equilibrium:
\begin{align}
	&\phi_o (b)=\psi (b),
	\label{BenchmarkB} \\
	&\psi (f)=0.
	\label{BenchmarkF}
\end{align}
From (\ref{RentFirm}), (\ref{RentHouseholds}), (\ref{BenchmarkB}), and (\ref{BenchmarkF}), the equilibrium wage and indirect utility are
\begin{equation*}
	w^*=p-2\tau b^2,\quad z^*=p-2\tau b^2-\kappa f,
\end{equation*}
where, when the parameter of FTF communication cost $ \tau $ and commuting cost $\kappa$ are high, the equilibrium wage $w^*$ and utility $z^*$ are low.
To satisfy the necessary conditions (\ref{BenchmarkB}) and (\ref{BenchmarkF}), equilibrium spatial conditions can be replaced by $ \phi_o(0)>\psi(0) $.
\footnote{The proof is in Appendix \ref{Conditions in benchmark case}}
Combining these conditions, the benchmark case is spatial equilibrium if and only if
\begin{equation}
	\frac{\kappa}{\tau} <\frac{2\bar{h}   b}{\bar{h}  +a_{s,o}}.
	\label{ConditionBenchmark}
\end{equation}
When commuting cost $ \kappa $ is sufficiently low and FTF communication cost $ \tau $ is sufficiently high, the benchmark case emerges.
Intuitively, when households accept long commuting and firms want to agglomerate in the business district for saving FTF communication costs, the monocentric city is equilibrium.

In the benchmark case, telework firms do not appear in the city.
From (\ref{CommutEQ}), (\ref{RentFirm}), and (\ref{FTFandWage}), firms bid rent functions are written by
\begin{equation}
	\phi_t (x)=\frac{1}{a_{s,t}}\left \{ p-\beta_t  (W(f)+MC_t)-(1-\beta_t) W(x)-\tau (1-\beta_t)   T(x) \right \}.
	\label{eq25}
\end{equation}
We state Lemma about the bid rent function of telework firms (\ref{eq25}) in the benchmark case.
\begin{lemma}
	As for bid rents of telework firms in the benchmark case, the following hold:
	(i) when telework cots are extremely high ($ MC_t \rightarrow \infty$), bid rent of telework firms extremely low ($ \phi_t(x) \rightarrow -\infty$),
	(ii) the bid rent of telework firms increase in parallel with the decline of telework costs $MC_t$ ($ \frac{d \phi_t (x)}{d MC_t}=-\frac{\beta_t}{a_{s,t}}<0 $, where $ \frac{\beta_t}{a_{s,t}}$ is constant).
	\label{lemma1}
\end{lemma} 
In the benchmark case, $MC_t$ is high enough and telework firms do not appear in the city.
When $ MC_t $ decreases sufficiently and telework firms' bid rents cross the market rent, telework firms appear in the city.
\footnote{Before telework firms appear in the city, the change of $MC_t$ is independent of endogenous variables $w^*$, $z^*$ and $\theta^*$.}
\subsection{Main Finding 1: Where in the city do firms employing teleworkers appear?}
As ICTs develop, firms employing teleworkers appear in cities because teleworking can be done for a small cost. Analytical analysis shows that the location of firms employing teleworkers are limited to near urban centers and near urban fringes.

\begin{figure}[H]
	\centering
	\begin{minipage}[b]{0.45\linewidth}
		\centering
		\includegraphics[keepaspectratio, scale=0.6]{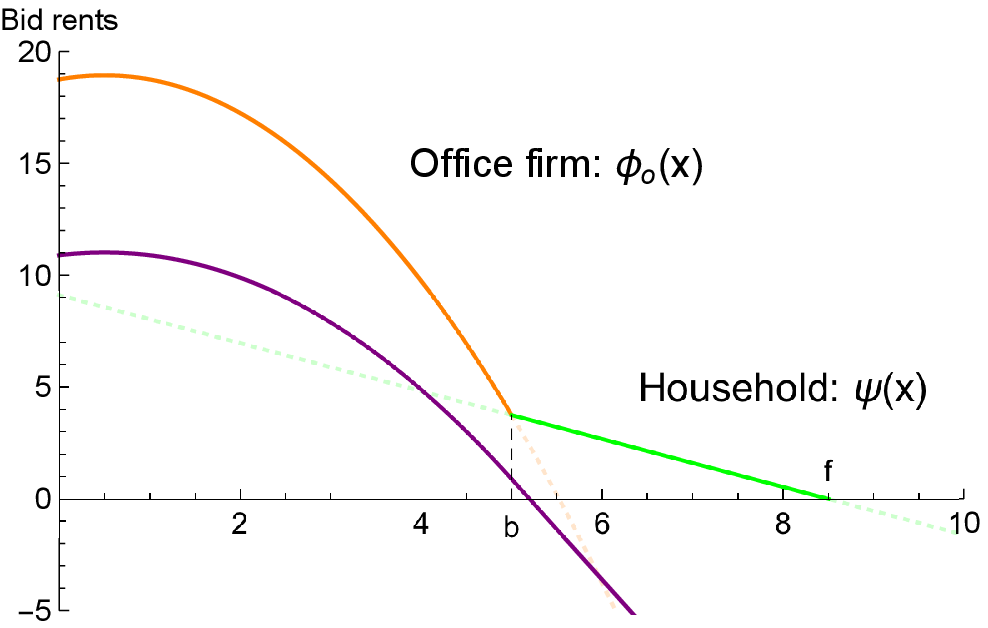}
	\end{minipage}
	\begin{minipage}[b]{0.45\linewidth}
		\centering
		\includegraphics[keepaspectratio, scale=0.6]{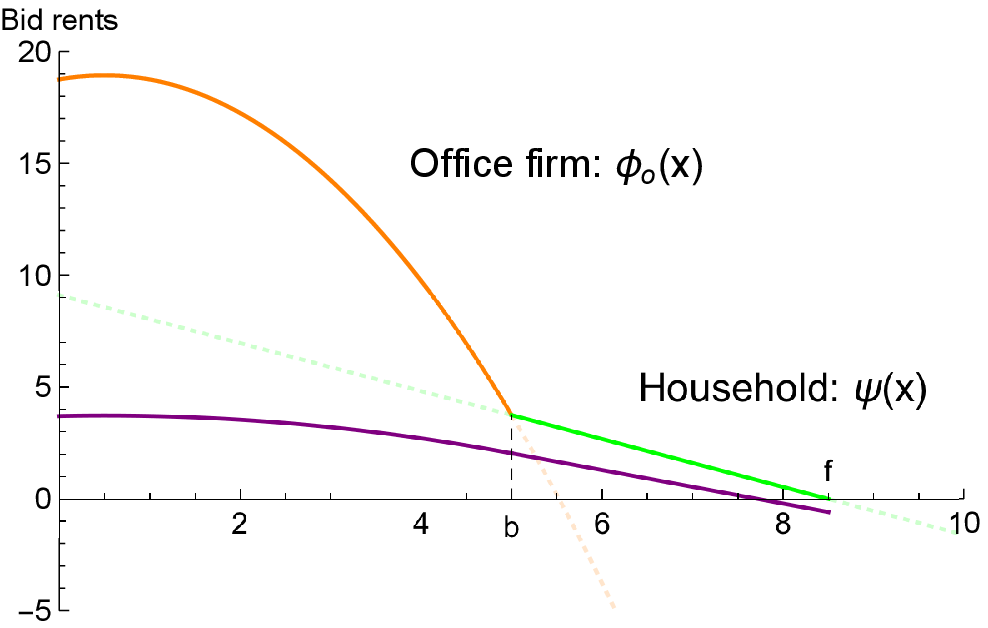}
	\end{minipage}
	\caption{\textbf{The location of telework firms depend on bid rent gradient} : If the bid rent gradient of telework firm is steeper (gentler) than that of households, telework firms are located at $ b $ ($ f $).
		These figures are made by benchmark parameter values, $ \beta_t= 0.4$ (left), and $ \beta_t= 0.9$ (right).}
	\label{LocFirstTele}
\end{figure}

The location of telework firms is determined by the slope of the firm's bid rent curve (Figure~\ref{LocFirstTele}).
In this analysis, we focus on the slope of the bid rent curve, i.e., commuting costs $\kappa$, FTF communication costs $\tau$, and teleworker ration $\beta_t$.
We compare the bid rent functions of the telework firms and office firms.
From  (\ref{RentFirm}) and (\ref{eq25}), bid rent gradient of type $ i $ firm in benchmark case is
\begin{equation}
	\phi_i '(x)=\frac{1-\beta_i}{a_{s,i}}  (\kappa-2\tau x),
	\label{FirmRentCurve}
\end{equation}
where the firm's bid rent function has the maximum value at $ y=\frac{\kappa}{2\tau} $ regardless of their types $i$.
As for the bid rent gradient, we make the basic assumption.
\begin{assumption}
	$\frac{a_{s,t}}{1-\beta_t}>a_{s,o}> $ is satisfied.
	\label{assumption1}
\end{assumption}
$ a_{s,i}/(1-\beta_i) $ represents a space per on-site worker.
\footnote{The previous study~\cite{miller2014workplace} have noted that newly established firms have larger office space per worker.
	This is because they contract for office space in anticipation of future growth, such as adding more employees.}
From (\ref{FirmRentCurve}) and Assumption \ref{assumption1}, the following holds.
\begin{lemma}
	(i) The bid rent gradient of office firms is always steeper than that of telework firms.
	(ii) Two types of firms have the same peak and are exogeneously given by $ y=\frac{\kappa}{2\tau} $, before telework firms appear in the city.
	\label{LemmaBidRentGradient}
\end{lemma}
From Lemma \ref{lemma1} and Lemma \ref{LemmaBidRentGradient}, locations where telework firms appear in the city are limited to $ x=b$ or $ f $.
\footnote{$ x=0 $ is excluded. That proof is in Appendix \ref{Proof:proposition1}.}
Intuitively, Figure~\ref{LocFirstTele} shows that whether the first telework firms appear at $ b $ or $ f $ depends on the bid rent gradient of telework firms and households.
If the bid rent gradient of telework firms is steeper than that of households, telework firms appear at $ b $.
Otherwise, they appear at $ f $.

\begin{figure} [H]
	\centering
	\includegraphics[scale=0.4]{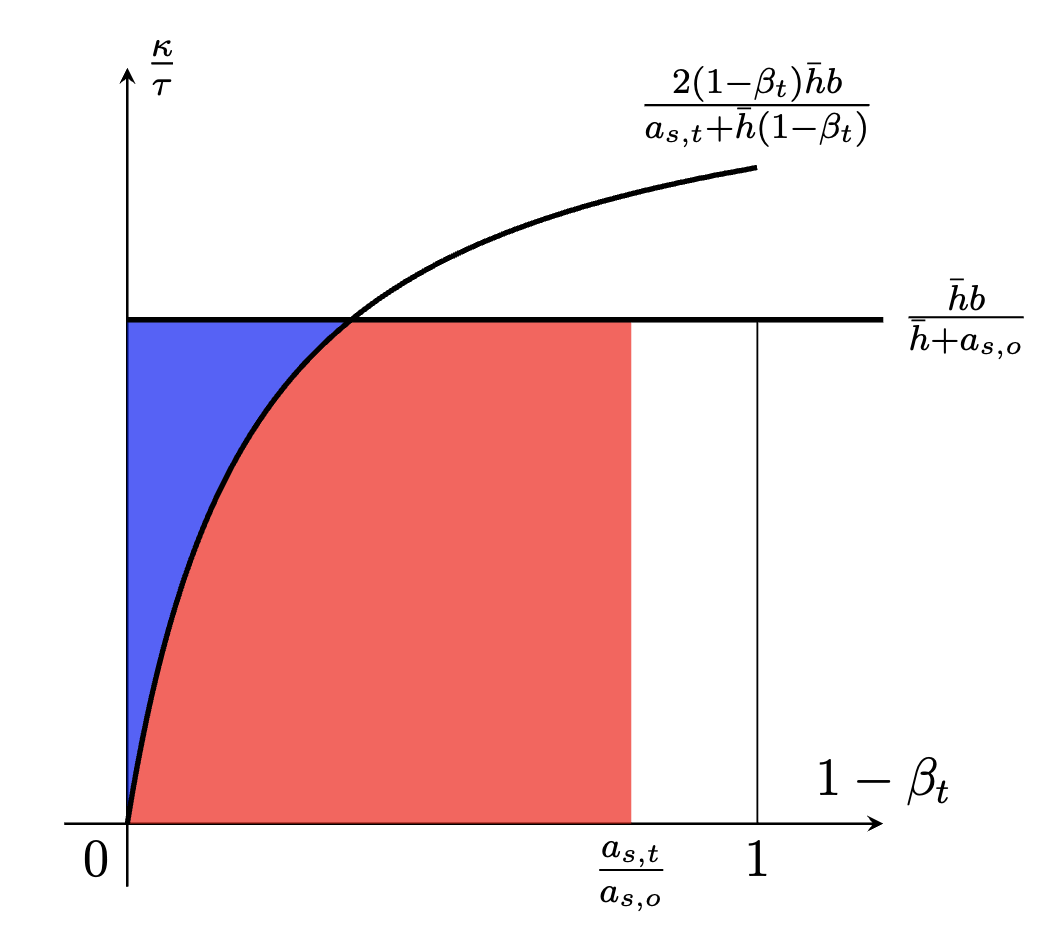}
	\caption{\textbf{Parametric conditions about where telework firms are located }: In the blue (red) area, telework firms appear at $ f $ ( $ b $ ). The location of telework firms are determined by three parameters: face-to-face communication $ \tau $, commuting cost $ \kappa $, and teleworker ratio $ \beta_t $.
		Colored areas are limited by two conditions $ 1-\beta_t<\frac{a_{s,t}}{a_{s,o}} $ and $\frac{\kappa}{\tau}<\frac{\bar{h}b}{\bar{h}+a_{s,o}}  $, which is from Assumption~\ref{assumption1} and (\ref{ConditionBenchmark}), respectably.}
	\label{FirstTele}
\end{figure}

\begin{proposition}
	When the telework costs ($ MC_t $) decrease, the following hold:
	(i) if $ \frac{\kappa}{\tau}<\frac{2(1-\beta_t)\bar{h}  b}{a_{s,t}+\bar{h}(1-\beta_t) }$, then telework firms appear at $ b $.
	(ii) if $ \frac{2(1-\beta_t)\bar{h}  b}{a_{s,t}+\bar{h}(1-\beta_t) }<\frac{\kappa}{\tau}<\frac{\bar{h}  b}{a_{s,o}+\bar{h}} $, then telework firms appear at $ f $.
	\label{proposition1}
\end{proposition}
The proof is presented in Appendix \ref{Proof:proposition1}.
Proposition~\ref{proposition1} is divided by commuting cost $ \kappa $, face-to-face communication costs $ \tau $, and teleworker ratio $ \beta_t $ and illustrated in Figure~\ref{FirstTele}.
If $\kappa/\tau$ and $ \beta_t $ is low (high), telework firms appear at \textit{CBD fringes} $ b $ (at \textit{urban fringes} $ f $).
Intuitively, when many households commute to their office, they accept commuting long distances, and firms want to agglomerate for saving their FTF communication cost, telework firms appear at \textit{CBD fringes} $b$.

\subsection{Main Finding 2: The Change of location equilibria and the impact of telework on cities}
Consider the impact on urban economies when the cost of teleworking is further decreased.
From Proposition~\ref{proposition1}, telework firms are located around $ b $ or $ f $.
We conduct a comparative statics for each location equilibrium to show that the expansion of teleworking makes cities more compact and production more efficient.

\subsubsection{The new telework firm appears at  CBD fringes $ b $}
\begin{center}
	\begin{figure}[h]
		\centering
		\includegraphics[width=400pt]{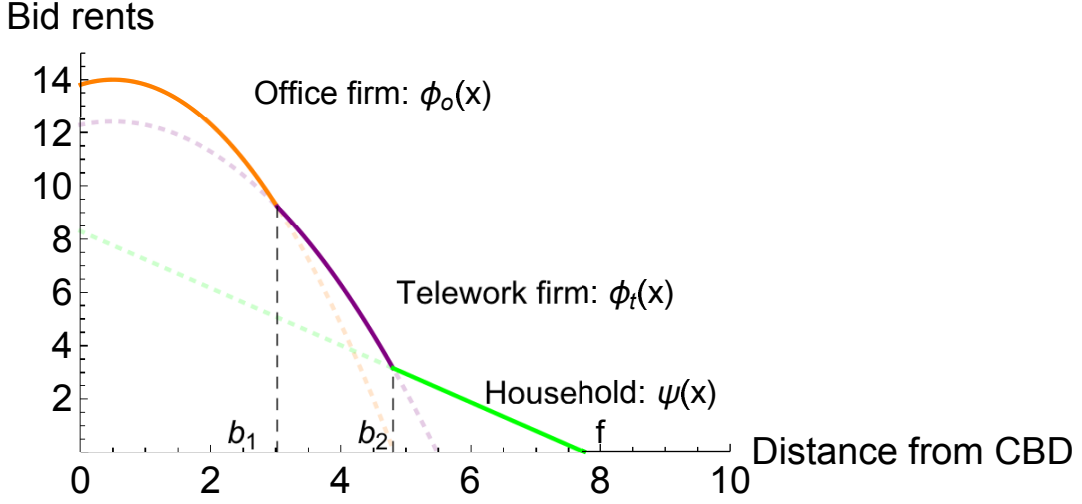}
		\caption{\textbf{An equilibrium land use pattern with telework: the first telework firms appear at $ b $ : }
			This numerical example is made by benchmark parameter values, $ \beta_t= 0.4$, and $ MC_t=6$.}
		\label{TeleB} 
	\end{figure}
\end{center}
If Proposition \ref{proposition1} holds (i), telework firms are located at $ b $.
When telework costs decrease further, telework expands in the city.
The expansion of teleworking makes cities more compact and production more efficient.

In the spatial equilibrium, telework firms are located form $ b_1 $ to $ b_2 $.
City boundaries are
\footnote{Detailed calculation is in Appendix~\ref{BordersB}.
	Indeed, $ b_1,  b_2 $, and $ f $  are functions of $ \theta $ but $ \theta $ is omitted for simplisity.}
\begin{equation}
	b_1= \frac{a_{s,o} M}{2}\theta,
	\quad
	b_2=b_1+ \frac{\bar{h}+a_{s,o}}{2} M(1-\theta),
	\quad
	f=b_2+\frac{1-(1-\theta)\beta_t}{2}M\bar{h}.
	\label{CityBoundariesB}
\end{equation}
FTF communication costs and the wage equation and are given by 
\begin{equation}
	W(x)=w-\kappa x,
	\quad
	T(x)=x^2+b_2^2,
	\quad
	x \in(0,b_2).
	\label{WageandFTFB}
\end{equation}
At city boundaries $b_1$, $b_2$, and $f$, bid rents satisfy the following nesessary conditions for spatial equilibrium:
\begin{align}
	&\phi_t(b_1)=\phi_o(b_1),
	\label{TeleB-B1}\\
	&\phi_t (b_2)=\psi (b_2),
	\label{TeleB-B2}\\
	&\psi (f)=0.
	\label{TeleB-F}
\end{align}
From (\ref{CityBoundariesB}) - (\ref{TeleB-F}), $\theta^\ast$, $w^\ast$, and $z^\ast$ cannot be solved in explicit form. \footnote{We can solve these equations only numerically. This result attaches the mathmatica file.}
However, we can examine the relationship between telework costs $MC_t$ and endogenous variable $\theta^\ast$, $w^\ast$, and $z^\ast$ by comparative statics.
\footnote{The derivative process of comparative statics is in Appendix~\ref{Calc:Comparative Statics B}} 
Table \ref{CSatB} summarizes the comparative statics analysis and we derive the following Proposition~\ref{Prop:ComparativeStaticsB}
\begin{table} [H]
	\begin{center}
		\caption{Results of comparative statics after telework firm appears around $ b $}
		\label{CSatB}
		\begin{tabular}{|c|c|c|c||c|c|c|c|c|c|}
			\hline
			& $ \theta^* $
			& $ w^* $
			& $ z^* $
			& $ b_1(\theta )$ 
			& $ b_2(\theta ) $ 
			& $f(\theta )$ 
			& $\phi_t(x) $ 
			& $\phi_o(x) $ 
			& $ \psi(x) $\\
			\hline
			$ MC_t $ & + & $ - $ & $ - $ & + & + & + & + & + & + \\
			\hline
		\end{tabular}
	\end{center}
\end{table}
\begin{proposition}
	If $ \frac{2(1-\beta_t)\bar{h}  b}{a_{s,t}+\bar{h}(1-\beta_t) }<\frac{\kappa}{\tau}<\frac{2\bar{h}  b}{\bar{h}a_L+a_{s,o}}$ holds and telework costs $ MC_t $ decrease further, telework firms are located from $ b_1 $ to $b_2$ and the following hold: 
	(i) The equilibrium welfare $z^*$, telework firms $ (1-\theta^*) $, and wages $w^*$ increase
	(ii) All city boundaries  ( $ b_1 $, $ b_2 $, and $ f $) shrink,
	(iii) All market rents ($ \phi_o $, $ \phi_t $, and $ \psi $) decrease.
	\label{Prop:ComparativeStaticsB}
\end{proposition}
By the decline of telework cost, competition between office and telework firms on land market are more severe.
Onsite firms replace to telework firms near \textit{CBD fringes} and teleworkers move out from the city. 
Hence, the city becomes compact and firms are more productive by telework.

\subsubsection{The new telework firm appears at urban fringes $ f $}
If Proposition \ref{proposition1} holds (ii), telework firms are located at $ f $.
As in the case appearing in $b$, the expansion of teleworking makes cities more compact and production more efficient.
However, the difference is that in location equilibrium, telework firms and households live mixed in the same location.

\begin{center}
	\begin{figure}[h]
		\centering
		\includegraphics[width=400pt]{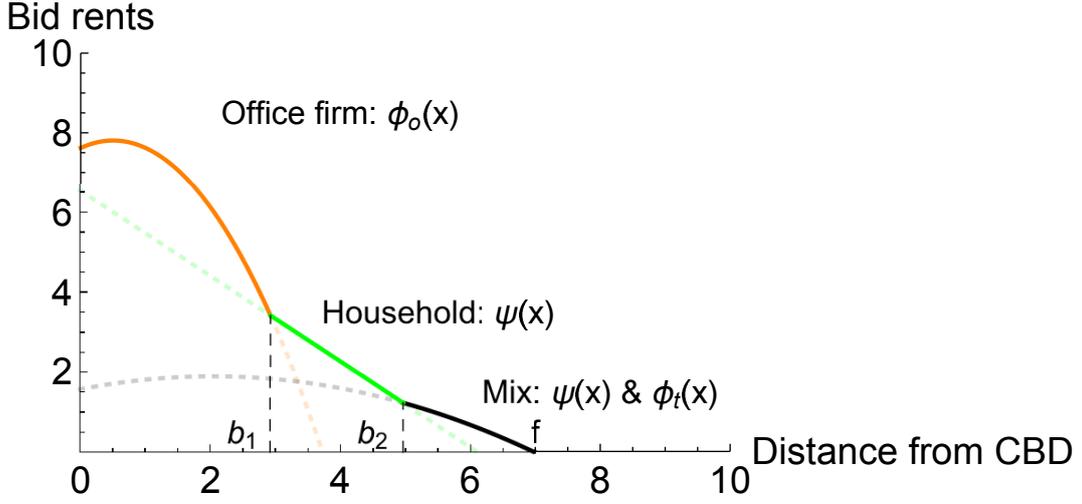}
		\caption{\textbf{An equilibrium land use pattern with telework: the first telework firms appear at $ f $.} This numerical example is made by benchmark parameter values, $ \beta_t= 0.9$, and $ MC_t=7$. In the mixed land use pattern (black line), households and telework firms colocate.}
		\label{TeleF} 
	\end{figure}
\end{center}

The spatial equilibrium is depicted in Figure~\ref{TeleF}.
Telework firms and households pay the same bid rent in $ y\in \left [ b_2,f \right ] $.
\begin{proposition}
	If $ \frac{2(1-\beta_t)\bar{h}  b}{a_{s,t}+\bar{h}(1-\beta_t) }<\frac{\kappa}{\tau}<\frac{2\bar{h}  b}{\bar{h}a_L+a_{s,o}}$ holds and telework costs $ MC_t $ decrease further, 
	telework firms colocate with households between $ b_2 $ and $ f $ in spatial equilibrium.
	\label{Mixb2-f}
\end{proposition}
The intuitive explanation is that workers who live at $r \in(b_1, b_2)$ commute to a telework firm at $y \in(b_2, f)$ can earn the same wage paying a lower rent by living near the telework firm. 
This violates the spatial equilibrium condition (\ref{EQ.Hosueholds}).
\footnote{ The proof is in Appendix~\ref{TeleHouseSegre}.}
From Figure~\ref{TeleF}, city boundaries are obtained by\footnote{Detailed calculation is in Appendix \ref{BordersF}.}
\begin{equation}
	b_1=\frac{a_{s,o} M}{2}\theta, 
	\quad 
	b_2=b_1+\frac{\bar{h} M}{2} \theta,
	\quad
	f=b_2+\frac{\bar{h}(1-\beta_t)+a_{s,t}}{2}(1-\theta) M.
	\label{CityBoundariesF}
\end{equation}
Necessary conditions of spatial equilibria are given by
\begin{align}
	&\phi_o(b_1)=\psi(b_1),
	\label{TeleF-b1}\\
	&\phi_t (x)=\psi (x),
	\quad x\in(b_2,f).
	\label{TeleF-b2f}
\end{align}
Firms'  FTF communication costs are given by
\begin{equation}
	T(y) = \left\{
	\begin{array}{ll}
		y^2+f^2+b_1^2-b_2^2 & y \in(0,b_1) \\
		y^2+f^2+2y(b_1-b_2) & y \in(b_2, f)
	\end{array}
	\right.
\end{equation}
From (\ref{CommutEQ}) and (\ref{TeleF-b2f}), the wage equation
\footnote{Teleworkers' wages are given by $W(f)=w_t =\frac{\bar{h}(p-\beta_t MC_t-\tau(1-\beta_t)T(f))+a_{s,t}z}{a_{s,t}+\bar{h}}$.}
is obtained by
\begin{equation}
	W(x)=\left\{
	\begin{array}{ll}
		w- \kappa x &x \in(0,b_2) \\
		\frac{\bar{h}(p-\beta_t(w_t+MC_t)-\tau(1-\beta_t)T(x))+a_{s,t}z}{a_{s,t}+(1-\beta_t)\bar{h}} & x \in(b_2, f)
	\end{array}
	\right.
	\label{wageTeleF}
\end{equation}
From (\ref{CityBoundariesF})-(\ref{wageTeleF}), results of comparative statics when telework firms appear at \textit{urban fringes} $f$, are summarized at Table~\ref{CSatF} and we derive the following Proposition~\ref{Prop:ComparativeStaticsF},
\footnote{ Detailed calculation is in Appendix \ref{Calc:Comparative Statics onlyF}}
\begin{table} [H]
	\centering
	\caption{Results of comparative statics and numerical calculations when telework firm appears at $ f $.}
	\label{CSatF}
	\begin{tabular}{|c|c|c|c||c|c|c|c|c|c|}
		\hline
		& $ \theta^* $
		& $ w^* $
		& $ z^* $
		& $ b_1(\theta )$ 
		& $ b_2(\theta ) $ 
		& $f(\theta )$ 
		& $\phi_t(x) $ 
		& $\phi_o(x) $ 
		& $ \psi(x) $\\
		\hline
		$ MC_t $ & + & $ - $ & $ - $ & + & + & + & + & + & + \\
		\hline
	\end{tabular}
\end{table}
\begin{proposition}
	If $ \frac{2(1-\beta_t)\bar{h}  b}{a_{s,t}+\bar{h}(1-\beta_t) }<\frac{\kappa}{\tau}<\frac{\bar{h}  b}{\bar{h}a_L+a_{s,o}}$ holds and telework costs $ MC_t $ decrease further, the following hold: 
	(i) the equilibrium welfare $z^*$, telework firms $ (1-\theta^*) $, and wages $w^*$ increase,
	(ii) all city boundaries ($b_1, b_2, f$) shrink,
	(iii) all market rents ($ \phi_o $, $ \phi_t $, and $ \psi $) decrease.
	\label{Prop:ComparativeStaticsF}
\end{proposition}
\noindent
These result are similar to Proposition~\ref{Prop:ComparativeStaticsB}.
When telework firms appear at $ f $, the city also become compact.
From (\ref{TeleF-b2f}) and (\ref{wageTeleF}), Figure~\ref{TeleF} is a spatial equilibrium if and only if 
\footnote{If $ \theta=1 $, this sufficient condition coincides with Proposition~\ref{proposition1}(ii) }
\begin{equation}
	\frac{2(1-\beta_t)\bar{h}}{a_{s,t}+\bar{h}(1-\beta_t)}(f-b_2+b_1)<\frac{\kappa}{\tau}<\frac{\bar{h}b_1}{a_{s,o}+\bar{h}}.
	\label{MixCondition}
\end{equation}
\section{Lot size and labor sifting in telework firms}\label{Section5}
The land and labor input of telework firms also affects their location and the land rent they can pay.
In this section, we show that the land and labor input of a telework firm changes its location and that a telework firm employing more on-site workers can pay higher land rent even if its telework costs are higher.

When telework firms appear in the city, firms at $ x $ satisfy zero profit condition, where any pairs of $ a_{s,t} $ and $ \beta_t $ are indifferent under the same telework costs $MC_t$.
Indifference curves are written by,
\begin{equation}
	a_{s,t}=\frac{1 }{\phi_t(x)}\{-\beta_t C(x)+p(x)\}.
	\label{Eq:IndiffCurve}
\end{equation}
where $ C(x)= w_t+ MC_t -w(x)-\tau T(x)$ and $ p(x) =p-w(x)-\tau T(x) $.
$ C(x) $ can interpret \textit{labor sifting cost} from one on-site workers to one teleworker.
Labor sifting cost $ C(x) $  plays an important role in determining the shape of the indifferent curve.
Here, we examine $C(x)$ focusing on location of telework firms at $b$ or $f$.
\footnote{Detailed calculation is in \ref{Teletype}.}
\begin{lemma}
	When telework firms are located at $ b $ and $ f $, the following hold:
	(i) When labor input shift from on-site worker to teleworker, labor sifting cost is positive at b and f: $ C(b), C(f) > 0 $.
	(ii) One onsite worker's cost at $ f $ is higher than that at $ b $:  $ w(b)+\tau T(b) < w(f)+\tau T(f)$.
	(iii) Labor input switching costs at $b$ is larger than that at $f$: $ C(f) < C(b) $.
	\label{LaborSwitchingCost}
\end{lemma}
A telework firm at $ f $ must incur more costs for hiring on-site workers than that at $b$ due to high FTF communication costs.
\footnote{The cost of teleworking ($w_t+MC_t$) is the same regardless of their location.}
Hence, if the firm at $b$ switches their labor input from on-site work to telework, the increase in labor costs is larger than that at $ f $. 
From Lemma~\ref{LaborSwitchingCost}, we compare the gradient of indifferent curves,
\begin{equation}
	-\frac{C(b)}{\phi_t(b)}<-\frac{C(f)}{\phi_t(f)}<0,
	\label{MRS}
\end{equation}
where the indifferent curve of telework firm at $ b $ is steeper than that at $ f $. 
In (\ref{MRS}),  $-C(x)/\phi_t(x)$ represents MRS (The Marginal Rate of Substitution) between land rent and labor shift.
When the telework firm consumes additional costs for one unit of land, this firm must give up hiring one unit of the teleworker and switch to one unit of the on-site worker. 

\begin{figure}[H]
	\centering
	\includegraphics[width=150mm]{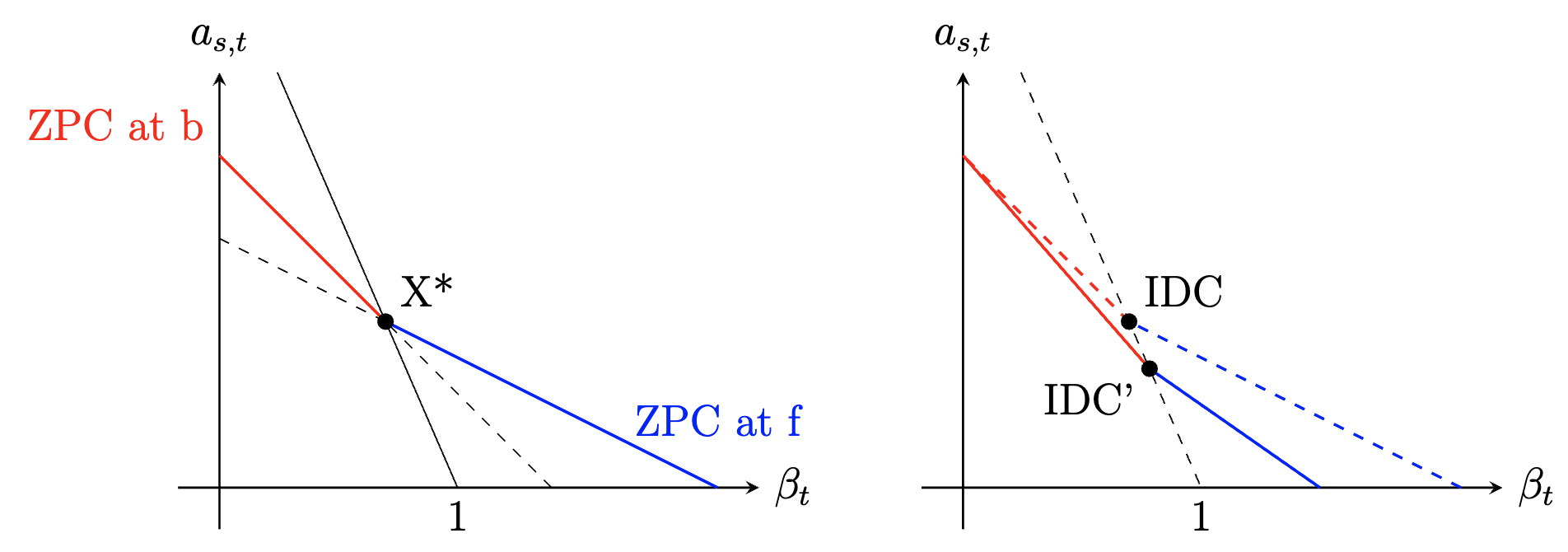}
	\caption{\textbf{Relationship between land productivity $a_{s,t}$ and teleworker ratio $\beta_t$}
		: The location of telework firms is also determined by land productivity and the teleworker ratio (left). 
		Furthermore, even if teleworking costs $MC_t$ are high, firms with a greater slope of the zero-profit condition appear in cities (right).}
	\label{MCtIndiff}
\end{figure}
We can illustrate Figure~\ref{MCtIndiff}(a) and (b) from (\ref{Eq:IndiffCurve}) and (\ref{MRS}).  
Figure~\ref{MCtIndiff}(a) indicates indifferent curves of firms locating at $ b $ and $ f $.
$ X^* $ shows the intersection of two curves.  
\footnote{$ X^* $ has $ (\beta^*_t  ,a_{s,t}^*) $ and such firms are indifferently located at $ b $ or $ f $.
	The interception of two indifferent curves $ X^* $ is obtained by
	\begin{equation*}
		(\beta^*_t, a^*_{s,t})=(\frac{\phi_t (f) p(b)-\phi_t (b) p(f)}{C(b) \phi_t(f) -C(f) \phi_t(b)},\frac{C(f) p(b)-C(b) p(f)}{C(f) \phi_t(b)-C(b) \phi_t(f) }),
	\end{equation*}
	where locating at whether $ b $ or $ f $ is indifferent on the same $ MC_t $.}
When telework costs $ MC_t $ are high, the indifference curve $ IDC $ moves to $ IDC$' in Figure~\ref{MCtIndiff}(b).
From Lemma~\ref{LaborSwitchingCost} and (\ref{MRS}), we obtain Proposition~\ref{TeleType&location} and  \ref{TeleType&MRS} about telework type (land productivity, MRS, telework costs $MC_t$) and their location,
\begin{proposition}
	Telework firms that have low land productivity (high $a_{s,t}$) and employ a large number of on-site workers (low $\beta_t$) appear in CBD fringes b. On the other hand, telework firms that have high land productivity (low $a_{s,t}$) and employ many teleworkers (high $\beta_t$) appear in urban fringes f.
	\label{TeleType&location}
\end{proposition}
Intuitively, telework firms at CBD fringe $b$ hire more on-site workers in the large office space and have similar input demands to office firms.
On the other hand, telework firms at $f$ hire more teleworkers in the small office spaces and have different input demands from office firms.
\begin{proposition}
	Telework firms with high $| -C(x)/\phi_t(x) |$ (MRS) and high land productivity (low $a_{s,t}$) can pay high rents and be located in the city, even if their telework costs $MC_t$ are high.
	\label{TeleType&MRS}
\end{proposition}
Intuitively, telework firms with a high MRS employ more on-site workers due to the higher labor shifting cost from the office to telework.
Such firms are not affected by high teleworking costs $MC_t$ and can pay higher rents to locate in cities with less office space (low $a_{s,t}$).

\section{Location externalities induced by telework firms}\label{Section6}
\begin{figure}
	\centering
	\includegraphics[width=150mm]{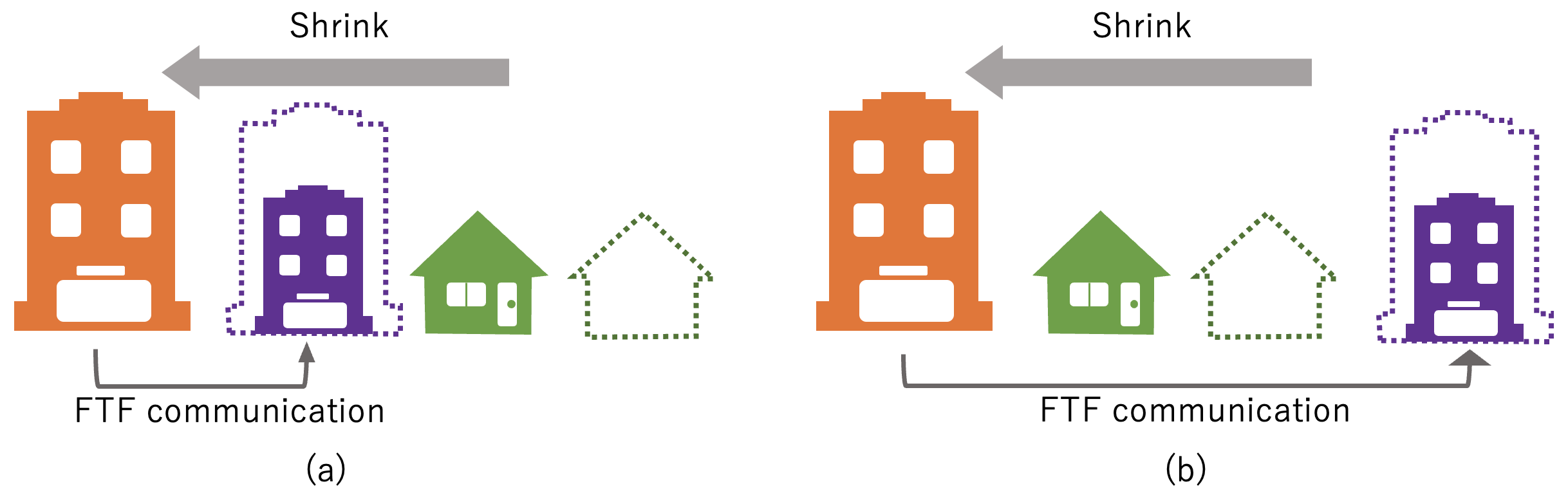}
	\caption{\textbf{Face-to-face (FTF) communications and externalities :}
		(a) All firms can save FTF communication costs and all households can save their commuting costs. 
		(b) All households can also save their commuting costs but FTF communication of firms is ambiguous because all firms near the city center must go to urban fringes for FTF communication.}
	\label{Externalities}
\end{figure}
This section investigates location externalities by telework firms.
The location of telework firms in the suburbs implies an increased economic burden on office firms located in the center.
The appearance of telework firms affects other office firms and households via commuting costs and FTF communication costs.
We define total urban costs $UC$ as the sum of total commuting costs $totalK$ and total FTF communication costs $totalT$ in the city and compare them before and after telework firms appear in the city, where the impact of these externalities vary depending on the location of the telework firm (Figure~\ref{Externalities}).

Before the telework firm emerges, the total urban costs $UC^{before}$ consists of total commuting $totalK^{before}$ and communication costs $totalT^{before}$ and is given as follows
\begin{align}
	\begin{aligned}
		&totalK^{before}=2\int_{b}^{f} \kappa y \,dy=\kappa(f^2-b^2),\\
		&totalT^{before}=\int_{-b}^{b} T (y)\,dy=\frac{3}{8} b^3.
	\end{aligned}
	\label{Total Costs}
\end{align}
The situation is divided into two cases, according to the location of telework firms at $ b $ and $ f $.
\subsection{Positive externalities at $ b $}
The total commuting cost and FTF communication cost after the telework firm appears at $ b $  are given by 
\begin{align}
	\begin{aligned}
		&total K^{after}(b)=2\int_{b-\Delta_b}^{f-\Delta_f} \kappa y\,dy=\kappa\{(f-\Delta_f)^2-(b-\Delta_b)^2\},\\
		&total T^{after}(b)=\int_{-b+\Delta_b}^{b-\Delta_b} T (y)\,dy=\frac{3}{8} (b-\Delta_b)^3,
	\end{aligned}
	\label{Total Costs at b}
\end{align}
where due to the emergence of telework firms \textit{CBD fringes} $b$ and \textit{urban fringes} $f$ shrink by $ \Delta_b $ and $ \Delta_f $, respectively.
Comparing ~(\ref{Total Costs}) and (\ref{Total Costs at b}), $ total K^{after}(b)<total K^{before} $ and $ total T^{after}(b)<total T^{before} $ always hold.
The total cost decreases by telework firm's emerging. (Figure~\ref{Externalities}-a).

\subsection{Negative and Positive externalities at $ f $}
On the other hand, negative externalities may occur when telework firms are located at $ f $.
The total commuting cost and FTF communication cost after the telework firm appears around $ f $  are given by 
\begin{align}
	\begin{aligned}
		&total K^{after}(f)=\kappa\{(f-\Delta_f)^2-(b-\Delta_b)^2\},\\
		&totalT^{after}(f)=\frac{3}{8}(b-\Delta_b)^3+4(f-\Delta_f)(b-\Delta_b).
	\end{aligned}
	\label{Total Costs at f}
\end{align}
Comparing ~(\ref{Total Costs}) and (\ref{Total Costs at f}), it is difficult to identify which is bigger, (\ref{Total Costs}) or (\ref{Total Costs at f}). 
However, when the number of telework firms are a few ($ \Delta_b $ and $ \Delta_f \simeq 0$) in (\ref{Total Costs at f}), the total urban cost is higher than (\ref{Total Costs}) due to telework firms.
From (\ref{Total Costs}) - (\ref{Total Costs at f}), we can get below Proposition~\ref{PropositionExternalities} about location externalities.
\begin{proposition}
	The location of telework firms generates positive or negative externalities on office firms and households.
	
	if $ \frac{\kappa}{\tau}<\frac{2(1-\beta_t)\bar{h}  b}{a_{s,t}+\bar{h}(1-\beta_t) }$, then telework firms appear at $ b $.
	(ii) if $ \frac{2(1-\beta_t)\bar{h}  b}{a_{s,t}+\bar{h}(1-\beta_t) }<\frac{\kappa}{\tau}<\frac{\bar{h}  b}{a_{s,o}+\bar{h}} $,
	
	(i)  If $ \frac{\kappa}{\tau}<\frac{2(1-\beta_t)\bar{h}  b}{a_{s,t}+\bar{h}(1-\beta_t) }$, telework firms appear at b and urban costs always decrease and positive externalities arise.
	(ii) If $ \frac{2(1-\beta_t)\bar{h}  b}{a_{s,t}+\bar{h}(1-\beta_t) }<\frac{\kappa}{\tau}<\frac{\bar{h}  b}{a_{s,o}+\bar{h}} $ and the number of telework firms is a few, telework firms appear at f and urban cost increase and negative externalities arise.
	\label{PropositionExternalities}
\end{proposition}
Intuitively, if telework firms are located at $f$, office firm must incur high FTF communication costs which exceed saved commuting costs (Figure~\ref{Externalities}-b).
The location of telework firms in the suburbs implies an increased economic burden on office firms located in the center.

However, we cannot remark social welfare from Proposition \ref{PropositionExternalities}, because our model has absentee landlords and is a partial equilibrium model.
Market rents absorb negative externalities and decrease in this model. 
This implies that income of absentee landlord decrease.

\section{Conclusion}\label{Section7}
In this study, we have revealed that the shift from office work to telework affects intra-urban location and economic activity.
The two main results obtained are as follows.
First, when telework costs $ MC_t $ decrease, the telework firms appear at CBD fringes $ b $ or urban fringes $ f $.
As for the parameter of commuting costs $\kappa$, face to face communication cost $\tau$, and teleworker ratio $\beta_t$, if $\kappa/ \tau $ and $ \beta_t$ high (low), telework firms are located at \textit{urban fringes} $f$ (\textit{CBD fringes} $b$).
This results can explain two empirical research (Maeng and Zorica, 2010~\cite{maeng2010relationship}; Liao, 2012~\cite{liao2012inshoring}) by one model.
Second, we revealed the mechanism of compact city by telework.
Expanding telework downsizes cities and increases firm productivity, wages, and welfare by comparative statics.
On the other hand, the total urban costs increase, when a few telework firms are located at $f$.
The location of telework firms in the suburbs implies an increased economic burden on office firms located in the center.
Our results support Kyriakopoulou and Picard (2021)~\cite{kyriakopoulou2022zoom} and suggest that telework enhances wages and welfare not only by telework promotion policy from the governments but also by the decline of telework costs from innovation.

Future studies could look more closely into the residential choice by teleworkers.
This study ignores the teleworkers' choice of residence.
In this model, teleworkers do not commute at all and have the same indirect utility wherever they live in the agricultural hinterland.
\footnote{Previous research assumed that onsite workers buy composite goods during their commuting, where they do not have to pay an additional cost for buying goods.
	So we need to assume that composite goods are freely traded and teleworkers buy them without trade costs.}
This assumption may be restrictive for analyzing city structure.
In the real world, teleworkers should choose their residence considering amenities and access to goods.

\appendix 

	\section{Calculation}
	\subsection{City boundaries in monocentric city equilibrium}	
	In the city, the ratio of office firms and telework firms are $ \theta:1-\theta $, $ \theta \in \left [ 0,1 \right ] $.
	This ratio is determined endogenously in spatial equilibrium.
	In telework firms, the ration of teleworker to onsite worker is exogenously given by $ \beta_t $.
	The total number of households in the city is  written by 
	\begin{equation}
		M-\beta_t(1-\theta)M
		\label{Households in City}
	\end{equation}
	Office firms need $ a_{s,o}  $ units of land and telework firms need $ a_{s,t}  $ units of land.
	\subsubsection{Benchmark case}\label{BordersBench}
	Firms are located from $ -b $ to $ b $. Households are located from $ b $ to $ f $ (from $ -f $ to $ -b $).
	The spatial equilibrium is symmetric.
	\begin{equation*}
		\int_{-b}^{b}\frac{1}{a_{s,o}}dx=M, \quad \int_{b}^{f}\frac{1}{\bar{h}}dx=\frac{M}{2}.
	\end{equation*}
	\subsubsection{Telework firms at b}\label{BordersB}
	Office firms are located from $ -b_1 $ to $ -b_1 $. Telework firms are located from $ b_1 $ to $ b_2 $ (from $ -b_2$ to $ -b_1 $). Households are located from $ b_2 $ to $ f $ (from $ -f $ to $ -b_2 $).
	From (\ref{Households in City}),
	\begin{equation*}
		\int_{-b_1}^{b_1}\frac{1}{a_{s,o}}dx=\theta M, \quad \int_{b_1}^{b_2}\frac{1}{a_{s,t}}dx=\frac{(1-\theta)M}{2},\quad
		\int_{b_2}^{f}\frac{1}{\bar{h}}dx=\frac{1-\beta_t(1-\theta)}{2}M.
	\end{equation*}
	\subsubsection{Telework firms at f}\label{BordersF}
	Households and telework firms colocate from $ b_2 $ to $ f $ (from $ -f $ to $ -b_2 $).
	In the mixed land use pattern, households do not commute and work in the residential place at $x$.
	From labor and land market clearing condition, $x \in (b_2,f)$ satisfies
	\begin{equation*}
		(1-\beta_t)m(x)=n(x),
		\quad
		a_{s,t}m(x)+\bar{h}n(x)=1.
	\end{equation*}
	The density of households $n(x)$ and telework firms $m(x)$ are given by
	\begin{equation*}
		m_t(x)=\frac{1}{\bar{h}(1-\beta_t)+a_{s,t}}, \quad n(x)=\frac{1-\beta_t}{\bar{h}(1-\beta_t)+a_{s,t}}.
	\end{equation*}
	where $\frac{1}{\bar{h}(1-\beta_t)+a_{s,t}}$ firms changed from office to telework firms.
	From (\ref{Households in City}),
	\begin{equation*}
		\int_{-b_1}^{b_1}\frac{1}{a_{s,o}}dx=\theta M, \quad 
		\int_{b_1}^{b_2}\frac{1}{\bar{h}}dx=\frac{\theta M}{2},\quad
		\int_{b_2}^{f}\frac{1}{\bar{h}(1-\beta_t)+a_{s,t}}dx=\frac{(1-\theta)}{2}M.
	\end{equation*}
	\subsection{The transition from the first telework firms appearing at $ b $ }
	From comparative statics, the relationship between $ \phi_o $, $ \phi_t $, and $ MC_t $ is given by,
	\begin{equation}
		\frac{d \phi_o(x)}{d MC_t}=\underset{+ }{\underbrace{\frac{d \phi_t(x)}{d MC_t}}}+(\underset{+ }{\underbrace{\frac{d \phi_t}{d b_1}-\frac{d \phi_t}{d b_2})\frac{d b_1}{d \theta}\cdot \frac{d \theta}{d MC_t}}}>0.
	\end{equation}
	where $ d \phi_o(x)/d MC_t>d \phi_t(x)/d MC_t $ always holds. 
	When telework costs ($ MC_t $) decrease in Figure~\ref{TeleB}, the decline of office firm's bid rent is bigger than that of telework firm.
	
	Now, when fixed costs for teleworkers decrease, bid rent of office work firms decline more than that telework firms. In Figure~\ref{TeleB}, the bid rent gradient of office firms always steeper than that of telework. And both bid rent functions monotonically increase at $ x\in [0,\kappa/2\tau] $.
	Thus, telework firms are located at another place, $ x=0 $.
	\begin{lemma}
		From Figure~\ref{TeleB}, when costs for teleworkers decrease continuously, telework firms are also located at $ x=0 $.
	\end{lemma}
	Firms' bid rents are linear form and steeper than households' bid rent in $ x>b_2 $.
	\footnote{From Proposition~\ref{proposition1}, if telework firms are located around $ b $, the bid rent gradient of telework firms is steeper than that of households.}
	No telework firms are located from $ b_2 $ to $ f $.  
	
	\section{Proof}
	\subsection{The range of $\kappa/\tau$ in benchmark case }\label{Conditions in benchmark case}
	\begin{proof}
		To satisfy the spatial equilibrium, the condition can be replaced by 
		$ \phi(0)>\psi(0) $ and $ \phi'(b)<\psi'(b) $.
		That is following holds
		\begin{align}
			&\phi_o(0)>\psi(0)
			\rightleftharpoons \quad
			\frac{\kappa}{\tau}<\frac{\bar{h}  b}{\bar{h}  +a_{s,o}},
			\label{eqB-1}\\
			&\phi'_o(b)<\psi'(b)
			\rightleftharpoons  \quad
			\frac{\kappa}{\tau}<\frac{2\bar{h} b}{\bar{h} +a_{s,o}}.
			\label{eqB-2}
		\end{align}
		From spatial equilibrium, $ \phi_t(0)>\psi(0) $ and $ \phi(b)=\psi(b) $.
		In conclusion, combining (\ref{eqB-1}) and (\ref{eqB-2}), I can calculate the range of $ \kappa/\tau$.
		This land use pattern is equilibrium if and only if (\ref{eqB-1})  holds.
	\end{proof}
	
	\subsection{The proof of Lemma \ref{LemmaBidRentGradient}}
	\begin{proof}
		From bid rent gradient (\ref{FirmRentCurve}),
		\begin{equation*}
			\phi_t '(y)=\frac{1-\beta_t}{a_{s,t}} (\kappa-2\tau y),\quad
			\phi_o '(y)=\frac{1}{a_{s,o}}(\kappa-2\tau y).
		\end{equation*}
		From Assumption (\ref{assumption1}), $ |\phi_o '(y)|>|\phi_t '(y)| $ always holds at any $ y $. 
	\end{proof}
	\subsection{The proof of Proposition \ref{proposition1}} \label{Proof:proposition1}
	\begin{proof}
		From (\ref{FirmRentCurve}) and Lemma \ref{LemmaBidRentGradient}, the location of telework firms will be limited to $ x=0, b$ or $ f $, then the following holds, 
		\begin{lemma}
			When $ MC_t $ decreases sufficiently, telework firms appear at $ x=0, b$ or $ f $
			\label{lemma3}
		\end{lemma}
		However, $ x=0 $ is excluded by mean rate of changes.
		The location of telework firms is divided into two parts according to two gradients (telework firm $ \phi_t'(x) $ and the household $ \psi'(x) $).
		If bid rent gradient of telework firm is steeper than that of households $ |\phi_t'(x)|> | \psi'(x)|$, telework firms appear in $x\in \left [0,b  \right ]$.
		If $ |\phi_t'(x)|< | \psi'(x)|$, it does in $x\in \left [b, f  \right ]$.
		It should be noticed that the function of FTF communication cost is different in the two cases.
		
		\subsubsection{The case of telework firms appear in $x\in \left [0,b  \right ]$}
		When telework firms are located in $x\in \left [0,b  \right ]$, FTF communication cost is written by
		\begin{equation*}
			T(x)=\int_{b}^{-b}\left | y-x \right |dy=x^2+b^2.
		\end{equation*}
		From Assumption \ref{assumption1}, $ \phi_o'(x)<\phi_t'(x) $ always holds.
		If the bid rent gradient of telework firm is steeper than that of household, $ \phi_t(x) $ touches the market rent $ R(x) $ at $ x\in \left [ 0,b \right ] $.
		This condition can be replaced by
		\begin{equation}
			\phi_t'(b)<\psi'(b) \leftrightharpoons \frac{\kappa}{\tau}<\frac{2(1-\beta_t)\bar{h}  b}{a_{s,t}+\bar{h}(1-\beta_t) }.
			\label{Tele at B}
		\end{equation}
		Next question is which place, $ 0 $ or $ b $, does the first telework firms appear.
		We answer this question by comparing average rate of changes between office firms and telework firms.
		\begin{align*}
			&\frac{\Delta \phi_o}{\Delta b}\equiv\frac{\phi_o(b)-\phi_o(0)}{b}=\frac{1}{a_{s,o}}(\kappa-\tau   b), \\
			&\frac{\Delta \phi_t}{\Delta b}\equiv\frac{\phi_t(b)-\phi_t(0)}{b}=\frac{1-\beta_t}{a_{s,o}}(\kappa-\tau   b),
		\end{align*}
		where $ \kappa-\tau   b<0 $. 
		$ \frac{\Delta \phi_o}{\Delta b}<\frac{\Delta \phi_t}{\Delta b} $ always holds and both are independent of $ MC_t $. 
		The average rate of changes in office firm is always steeper than that of telework firm.
		Thus, the bid rent of telework firms touches market rent from below at $ b $.
		\begin{lemma}
			$\frac{\kappa}{\tau}<\frac{2(1-\beta_t)\bar{h}  b}{a_{s,t}+\bar{h}(1-\beta_t) }$, the first telework firms appear at $ b $.
			\label{lemma: Tele at B}
		\end{lemma}
		
		\subsubsection{The case of telework firms appear in $x\in \left (b,f  \right ]$}
		When telework firms appear in $x\in \left (b,f  \right ]$, the bid rent gradient of households is steeper than that of telework firms.
		FTF communication cost of telework firms is written by,
		\begin{equation*}
			T(x)=\int_{-b}^{b}|y-x|dy+2f=2(xb+f),
		\end{equation*}
		Both bid rent gradient of telework firms $ \phi'_t (x) $ and households $ \psi' (x) $ are linear form and given by
		\begin{equation*}
			\phi'_t (x)=\frac{\beta_i}{a_{s,t}} (\kappa-2 \tau b) , \quad  \psi' (x)=-\frac{\kappa}{\bar{h}}.
		\end{equation*}
		If the bid rent gradient of households is steeper than that of telework firms and both are linear form, $ \phi_t(x) $ crosses $ R(x)=\phi_t(x) $ only at $ f $. This condition can be replaced by
		\begin{equation}
			\phi'_o(x)<\psi'(x)<\phi'_t(x) \leftrightharpoons \frac{2(1-\beta_t)\bar{h}  b}{a_{s,t}+\bar{h}(1-\beta_t) }<\frac{\kappa}{\tau}<\frac{\bar{h}  b}{\bar{h}a_L+a_{s,o}}.
			\label{Tele at F}
		\end{equation}
		From Lemma \ref{lemma: Tele at B} and (\ref{Tele at F}), we can get Proposition~\ref{proposition1}.
	\end{proof}
	\subsection{The proof of Proposition~\ref{Mixb2-f}}\label{TeleHouseSegre}
	In equilibrium, telework firms around \textit{urban fringes} cannot be located independently.
	In spatial equilibrium, telework firms must hire at least one onsite worker.
	A household who lives at $r \in (b_1, b_2)$ and commutes to telework firms $y \in (b_2, f)$ toward urban fringes has indirect utility $ z(r,y)=W(y)-\phi(r)\bar{h}$.
	However, if she lives at $ y $, she can get the same wage $ W(y) $ and live by cheaper land rent $ \phi(y)<\phi(r) $.
	Thus, she can get  the higher indirect utility $ z(y,y)>z(r,y) $.
	This violates the spatial equilibrium condition~\ref{EQ.Hosueholds}.
	We can get below Lemma about households' commuting pattern.
	\begin{lemma}
		Households living at $ r $ do not commute to firms outside of their residence at $ y $, where $ r, y \in(0,f) $ always satisfies $ r \leq y $.
		\label{OutCommu}
	\end{lemma}
	Thus, households and telework firms are not segregated in spatial equilibrium and mixed land use pattern appears around \textit{urban fringes}.
	
	\subsection{Proof of Lemma \ref{LaborSwitchingCost}}\label{Teletype}
	We rewrite (\ref{Eq:IndiffCurve}) as
	\begin{equation*}
		p(x)-a_{s,t}\phi_t(x)=\beta_t C(x)
	\end{equation*}
	where the LHS is obviously positive ($p-w(x)-\tau T(x)-a_{s,t}\phi_t(x)>0$).
	Labor shift cost is positive and $ C(b), C(f) > 0 $ holds.
	If telework firms shift their labor input from on-site workers to teleworkers, they have to incur additional costs.
	Two indifferent curves intersect at $(\beta^*, a^*_{s,t})$.
	\begin{displaymath}
		\left\{
		\begin{array}{ll}
			p(b)-a^*_{s,t} \phi_t(b)=\beta^* C(b) \\
			p(f)-a^*_{s,t} \phi_t(f)=\beta^*C(f)
		\end{array}
		\right.
	\end{displaymath}
	And we get
	\begin{equation*}
		(1-\beta^*)(w(f)+\tau T(f)-w(f)-\tau T(f))=a^*_{s,t}(\phi_t(b)-\phi_t(f))
	\end{equation*}
	where, comparing bid rents,$\phi_t(b)-\phi_t(f)>0$ is obviously satisfied and $w(f)+\tau T(f)-w(f)-\tau T(f)>0$ holds.
	One unit of on-site worker at $b$ is more expensive than that at $f$ since face-to-face communication costs are burdensome for telework firms at $f$. 
	From (i) and (ii), $C(b)>C(f)$ holds and telework firms at $b$ must incur high labor switch costs those at $f$.
	
	\section{Comparative statics}
	Using comparative statics to spatial equilibrium conditions, we reveal the relationship between endogenous variables and costs for telework. 
	\subsection{Telework firms are located around $ b $}\label{Calc:Comparative Statics B}
	Explicit forms of bid rent functions are
	\begin{align}
		\begin{aligned}
			&\phi_o(x)=\frac{1}{a_{s,o}}\{p- (w-\kappa x)-\tau   (x^2+b_2^2)\},\\
			&\phi_t(x)=\frac{1}{a_{s,t}}\{p-\beta_t  (w_t+MC_t)-(1-\beta_t)  (w-\kappa x)-\tau (1-\beta_t)   (x^2+b_2^2)\},\\
			&\psi(x)=\frac{1}{\bar h}(w-\kappa x-z).
			\label{MarketRents-B}
		\end{aligned}
	\end{align}
	We totally differentiate (\ref{TeleB-B1}) by costs for telework, $ MC_t $,
	\begin{equation*}
		\begin{split}
			&(\frac{\partial \phi_o}{\partial b_1}\cdot\frac{\partial b_1}{\partial \theta}+\frac{\partial \phi_o}{\partial b_2}\cdot\frac{\partial b_2}{\partial \theta})\,d\theta+\frac{\partial \phi_o}{\partial w}\,dw\\
			&\quad=(\frac{\partial \phi_t}{\partial b_1}\cdot\frac{\partial b_1}{\partial \theta}+\frac{\partial \phi_t}{\partial b_2}\cdot\frac{\partial b_2}{\partial \theta}+\frac{\partial \phi_t}{\partial f}\cdot\frac{\partial f}{\partial \theta})\,d\theta+\frac{\partial \phi_t}{\partial w}\,dw+\frac{\partial \phi_t}{\partial MC_t}\,dMC_t.
		\end{split} 
	\end{equation*}
	From total differentiation, we have
	\begin{equation}
		\{\underset{-A_{11}< 0}{\underbrace{ \left (  \frac{\partial \phi_o}{\partial b_1}-\frac{\partial \phi_t}{\partial b_1}\right )\frac{\partial b_1}{\partial \theta}+\left ( \frac{\partial \phi_o}{\partial b_2}-\frac{\partial \phi_t}{\partial b_2} \right )\frac{\partial b_2}{\partial \theta}-\frac{\partial \phi_t}{\partial f}\frac{\partial  f}{\partial \theta} }}\}
		\,d\theta+(\underset{A_{12}>0}{\underbrace{\frac{\partial \phi_o}{\partial w}-\frac{\partial \phi_t}{\partial w}}})\,dw=\frac{\partial \phi_t}{\partial MC_t}\,dMC_t,
		\label{TotalDiff-b1}
	\end{equation}
	where $  \frac{\partial \phi_o}{\partial b_i}-\frac{\partial \phi_t}{\partial b_i}<0 $ from Lemma \ref{LemmaBidRentGradient}. $ \frac{\partial \phi_o}{\partial w}-\frac{\partial \phi_t}{\partial w} $,  $\frac{\partial \phi_t}{\partial f}$ and $\frac{\partial  f}{\partial \theta}$ are obtained by
	\begin{align*}
		&\frac{\partial \phi_t}{\partial f}=\frac{1}{a_{s,t}}\beta_t   \kappa >0\\
		&\frac{\partial  f}{\partial \theta}=(a_{s,o}-a_{s,t}) \frac{M}{2}>0\\
		&\frac{\partial \phi_o}{\partial w}-\frac{\partial \phi_t}{\partial w}=-\frac{1 }{a_{s,o}}+\frac{1 }{a_{s,t}}>0
	\end{align*}
	We totally differentiate (\ref{TeleB-B2}) and (\ref{TeleB-F}) in the same way and get
	\begin{align}
		&
		\begin{split}
			(\frac{\partial \phi_t}{\partial b_2}\cdot\frac{\partial b_2}{\partial \theta}+\frac{\partial \phi_t}{\partial f}\cdot\frac{\partial f}{\partial \theta})\,d\theta+&\frac{\partial \phi_t}{\partial w}\,dw+\frac{\partial \phi_t}{\partial MC_t}\,dMC_t
			\\&=\frac{\partial \psi}{\partial w}\,dw+\frac{\partial \psi}{\partial b_2}\cdot\frac{\partial b_2}{\partial \theta}\, d\theta+\frac{\partial \psi}{\partial z}\,dz
		\end{split}
		\label{TotalDiff-b2} \\
		& \frac{\partial \psi}{\partial w}\,dw+\frac{\partial \psi}{\partial f}\cdot\frac{\partial f}{\partial \theta}\, d\theta+\frac{\partial \psi}{\partial z}\,dz=0.
		\label{TotalDiff-f}
	\end{align}
	Combining (\ref{TotalDiff-b2}) and (\ref{TotalDiff-f}),
	\begin{equation}
		\{\underset{-A_{21}<0}{\underbrace{\left ( \frac{\partial \phi_t}{\partial b_2}-\frac{\partial \psi}{\partial b_2} \right )\frac{\partial b_2}{\partial \theta}+\left ( \frac{\partial \phi_t}{\partial f}-\frac{\partial \psi}{\partial f} \right )\frac{\partial f}{\partial \theta}}}\}\,d\theta+\frac{\partial \phi_t}{\partial w}\,dw=-\frac{\partial \phi_t}{\partial MC_t}\,dMC_t
		\label{TotalDiff-b2f}
	\end{equation}
	From (\ref{TotalDiff-b1}) and (\ref{TotalDiff-b2f}),
	\begin{align}
		\begin{aligned}
			\begin{bmatrix}
				-A_{11} & A_{12} \\ 
				-A_{21} & \frac{\partial \phi_t}{\partial w}
			\end{bmatrix}
			\begin{bmatrix}
				\frac{d \theta}{d MC_t}\\
				\frac{d w}{d MC_t}
			\end{bmatrix}
			&=
			\begin{bmatrix}
				\frac{\partial \phi_t}{\partial MC_t}\\
				-\frac{\partial \phi_t}{\partial MC_t}
			\end{bmatrix}\\
			\begin{bmatrix}
				\frac{d \theta}{d MC_t}\\
				\frac{d w}{d MC_t}
			\end{bmatrix}
			&=
			\begin{bmatrix}
				(\frac{\partial \phi_t}{\partial w}+A_{12})\frac{\partial \phi_t}{\partial MC_t}>0\\
				(A_{21}+A_{11})\frac{\partial \phi_t}{\partial MC_t}<0
			\end{bmatrix},
			\label{Results_w&theta-B}
		\end{aligned}
	\end{align} 
	where $ \frac{d \theta}{d MC_t} $ is positive and $ \frac{d w}{d MC_t} $ is negative.
	And from (\ref{TotalDiff-f}) and (\ref{Results_w&theta-B}), $\frac{d z}{d MC_t} $ is positive:
	\begin{equation}
		\frac{\partial z}{\partial MC_t}=\frac{\partial w}{\partial MC_t}-\kappa \frac{\partial f}{\partial MC_t} <0.
		\label{Results_z-B}
	\end{equation}
	When telework costs $MC_t$ decrease, the ratio of telework firm, wages, and welfare increases.
	Households are better off by the reduction in telework costs $MC_t$.
	From (\ref{Results_w&theta-B}) and (\ref{Results_z-B}), the effects on city boundaries by the decline of telework costs are written in 
	\begin{align}
		\begin{aligned}
			&\frac{\partial b_1}{\partial MC_t}=a_{s,o}\frac{\partial \theta}{\partial MC_t}\frac{ M}{2}>0,\\
			&\frac{\partial b_2}{\partial MC_t}=(a_{s,o}-a_{s,t})\frac{\partial \theta^\ast}{\partial MC_t}\frac{ M}{2}>0,\\
			&\frac{\partial f}{\partial MC_t}=\{\bar{h}\beta_t +(a_{s,o}-a_{s,t})\}\frac{\partial \theta}{\partial MC_t}\frac{ M}{2}>0,
		\end{aligned}
		\label{Results_CityBorders-B}
	\end{align}
	where the decline of telework costs $MC_t$ makes all city boundaries shrink.
	From (\ref{MarketRents-B}) and (\ref{Results_w&theta-B})-(\ref{Results_CityBorders-B}), the alternation of bid rent functions is obtained by
	\begin{align}
		\begin{aligned}
			&\frac{\partial \psi}{\partial MC_t}=\bar{h} \kappa \frac{\partial f}{\partial \theta} \frac{\partial \theta}{\partial MC_t}>0,\\
			&\frac{\partial \phi_t(x)}{\partial MC_t}=-\kappa (\frac{\partial b_2}{\partial \theta}-\frac{\partial f}{\partial \theta})\frac{\partial \theta}{\partial MC_t}>0,\\
			&\frac{\partial \phi_o(x)}{\partial MC_t}=\frac{\partial \phi_t(x)}{\partial MC_t}+(\frac{\partial \phi_t}{\partial b_1}+\frac{\partial \phi_t}{\partial b_2})\frac{\partial b_1}{\partial \theta}\frac{\partial \theta}{\partial MC_t}>0,
		\end{aligned}
		\label{Results_Rents-B}
	\end{align}
	where market rents decrease in city by reduction for teleworks' cost.
	Results of (\ref{Results_w&theta-B})-(\ref{Results_Rents-B}) are summarized in Table~\ref{CSatB}.
	
	\subsection{Telework firms are located at only $f$.}
	\label{Calc:Comparative Statics onlyF}
	Necessary conditions of spatial equilibrium are given by
	\begin{align}
		&\phi_o(b)=\psi(b)
		\label{TeleF-B}\\
		&\phi_t (f)=\psi(f)=0
		\label{TeleF-F}
	\end{align}
	From \ref{BordersF}, $\frac{1}{\bar{h}(1-\beta_t)+a_{s,t}}$ firms changed from office to telework firms and $\frac{\beta_t}{\bar{h}(1-\beta_t)+a_{s,t}}$ households became telework in the right hand side.
	\textit{CBD fringes} $b$ and \textit{Urban fringes} $f$ are given by
	\begin{align*}
		b^*&=\underbrace{a_{s,o}\frac{M}{2}}_{Benchmark\;case's\;b}-\underbrace{\frac{a_{s,o}}{\bar{h}(1-\beta_t)+a_{s,t}}}_{Number\;of\;onsite\;firms\;decreased}\\
		f^*&=\underbrace{benchmark case's f}_{(a_{s,o}+\bar{h})\frac{M}{2}}
		-\underbrace{\frac{a_{s,o}}{\bar{h}(1-\beta_t)+a_{s,t}}+\frac{a_{s,t}}{\bar{h}(1-\beta_t)+a_{s,t}}}_{Change of\;land\;use pattern}
		-\underbrace{\frac{\bar{h}\beta_t}{\bar{h}(1-\beta_t)+a_{s,t}}}_{Teleworkers'\;moving\;out}\\
		&=(a_{s,o}+\bar{h})\frac{M}{2}-\frac{a_{s,o}-a_{s,t}+\bar{h}\beta_t}{\bar{h}(1-\beta_t)+a_{s,t}}
	\end{align*}
	We consider the case where teleworking firms are located only at \textit{urban fringes} $f$.
	FTF communication costs are written by
	\begin{equation*}
		T(x) = \left\{
		\begin{array}{ll}
			2(x^2+f) & x \in (0,b_1)\\
			2f(b_1+1) & x = f
		\end{array}
		\right.
	\end{equation*}
	From (\ref{TeleF-B}) and (\ref{TeleF-F}), endogenous variables are obtained by
	\begin{align}
		\begin{aligned}
			&w^*=p-2\tau (b_1^2+f)-\frac{\kappa a_{s,o}}{\bar{h}},
			\quad
			z^*=w_t^*=p-2\tau (b_1^2+f)-\kappa (\frac{a_{s,o}}{\bar{h}}+f),\\
			& \theta^*=1-\frac{2}{\left\{ a_{s,t}+\bar{h}(1-\beta_t) \right\}M}.
		\end{aligned}
		\label{TeleatFonly}
	\end{align}
	From (\ref{TeleatFonly}), the bid rent function of telework firms are 
	\begin{equation*}
		\phi_t(x; \theta=\theta^*)=\frac{1}{a_{s,t}} \left\{ p-(1-\beta_t)(w^*-\kappa x)-\beta_t (w_t+MC_t)-2(1-\beta_t)f(b_1+1)\right\},
	\end{equation*}
	where $\frac{\partial \phi_t(f)}{\partial MC_t}=\frac{-\beta_t}{a_{s,t}}<0$.
	When telework costs $MC_t$ decrease, the bid rent of telework firms increases.
	This result indicates that the number of telework firms increases.

\end{document}